# Raman Spectroscopy and Machine Learning-based Optical Sensor for Rapid Tuberculosis Diagnosis via Sputum

Ubaid Ullah[1], Zarfishan Tahir[2], Obaidullah Qazi[2], Shaper Mirza[3], and M. Imran Cheema[1*]

[1] Department of Electrical Engineering, Syed Babar Ali School of Science and Engineering, Lahore University of Management Sciences, Lahore, Pakistan

[2] Institute of Public Health, Lahore, Pakistan

[3] Department of Biology, Syed Babar Ali School of Science and Engineering, Lahore University of Management Sciences, Lahore, Pakistan

[*]imran.cheema@lums.edu.pk

**ABSTRACT**

Tuberculosis (TB) is a contagious disease that causes 1.5 million deaths per year globally. Early diagnosis of TB patients is critical to control its spread. However, standard TB diagnostic tests such as sputum culture take days to weeks to produce results. Here, we demonstrate a quick, portable, easy-to-use, and non-invasive optical sensor based on sputum samples for TB detection. The probe uses Raman spectroscopy to detect TB in a patient's sputum supernatant. We deploy a machine-learning algorithm, principal component analysis (PCA), on the acquired Raman data to enhance the detection sensitivity and specificity. On testing 112 potential TB patients, our results show that the developed probe's accuracy is 100% for true-positive and 93.4% for true-negative. Moreover, the probe correctly identifies patients on TB medication. We anticipate that our work will lead to a viable and rapid TB diagnostic platform.

## 1. Introduction

TB is one of the deadliest infectious diseases impacting in one-quarter of the world population developing the infection. TB's impact is severe in low and middle-income nations, causing 4000 deaths daily. The worrying TB statistics are due to its high infection rate and no easy access to diagnostic facilities, especially in developing countries [1]. Given the highly contagious nature of TB, early and onsite detection can reduce disease transmission and improve mortality rate by allowing the administration to control its spread through proper medication and isolation of patients. The gold-standard methodology for TB detection is sputum culture; however, the test takes days to weeks to produce results [2, 3]. During this time, the TB-positive person is likely to infect close contacts, thereby increasing transmission in the community. Moreover, the culture test needs trained laboratory personnel for various processing steps. Finally, patients are asked for multiple visits to diagnostic centers [4], which is often difficult for people living in remote areas with limited resources [5].

Besides high throughput, an ideal TB sensor should provide high sensitivity and specificity, have a cheap operational cost, should not require specialized personnel, be portable, and the waste should be environmentally friendly [6]. Achieving these specifications for a TB sensor has been the focus of various research works based on immunochromatography [7], polymerase-chain-reaction (PCR) [8], piezoelectricity [9], electrochemical [10], and surface plasmon resonance [11]. However, all of these techniques suffer from one or more fundamental limitations such as low sensitivity and specificity, invasiveness, non-portability, and the requirement of a specialist to operate. In contrast, optical sensors, specifically those based on Raman spectroscopy, have the potential to overcome these limitations.

In Raman spectroscopy, light passes through a sample (solid, liquid, and gas), interacts, and excites different molecular vibrations at specific frequencies. Every functional group excites unique molecular vibrations specific to material structure (e.g., C-C vs. C=C) and the adjacent groups. These unique signatures can accurately identify materials. Raman spectroscopy has been widely applied for explosive identification and, most notably, was the key method used to identify water on Mars [12]. In the last decade, Raman spectroscopy in various configurations has been thoroughly investigated to detect multiple diseases, including cancer, endoscopy, and skin disorders [13-15]. One of the successful implementations of Raman systems in various forms is the diagnosis of different types of cancer with clinical trials on hundreds of patients [14, 16, 17]. Companies now exist that provide commercial Raman systems for various disease diagnoses, e.g., Kaiser optical system Inc., USA, develops Raman systems for multiple applications, including cancer, skin, bone, eyes, and dental diagnosis [18].

Some researchers have utilized Raman spectroscopy to detect TB in blood and TB meningitis in cerebrospinal fluid [19, 20]. Even though these approaches accelerate TB diagnosis, their invasive nature limits their applications. A subsequent study that employed Raman spectroscopy of individual mycobacterium cells for TB detection offered species-level information [21]. They achieved an accuracy of 94%; however, the complex sample preparation process necessitates a specialist's need. Furthermore, human sampling was missing in the study.

It is well-researched that the sputum of a person with TB has a range of biomarkers, including esters and proteins [22-24]. We hypothesize that the presence of these compounds minutely changes the Raman signature of the sputum's supernatant sample even after filtering Mycobacterium tuberculosis (Mtb) bacteria cells. The change in the Raman signature is weak to identify TB biomarkers but is sufficient to probe TB. However, its weak nature makes it hard to differentiate the probe signal visually. Therefore, we utilized a machine learning algorithm, principal component analysis (PCA), for TB-positive and TB-negative samples classification. We tested three groups of patients: TB-, first-time TB+, and TB+ under treatment. Our developed sensor classified TB infected and non-infected patients with 100% true-positive and 93.4% true-negative accuracy.

Now we describe the rest of the paper. Firstly, we provide patient information, sample collection, and preparation methods. Secondly, we discuss the experimental setup and its working principle, Raman spectrum collection, and the post-processing algorithm, Principal Component Analysis (PCA). Thirdly, we present experimental results and their significance, and finally, we provide our main findings and concluding remarks.

## 2. Patients and Methods

The project was approved by the Institutional Review Board (protocol no. LUMS/IRB-2016-06-03) of Lahore University of Management Sciences. All experiments were performed in accordance with guidelines for Good Laboratory Practices (CDC). We used archived material of patients, and consent was obtained on the original samples. Sputum samples were collected from 112 patients in the age group of 14-58 years old. Samples were collected from different provincial TB centers and analyzed in a control lab environment at the Institute of Public Health, Lahore, Pakistan, using our proposed TB diagnostic probe. We evaluated the probe's performance using the start-of-art Mycobacteria Growth Indicator Tube (MGIT) culture test. Our study included 45 TB-, 18 first-time TB+, and 49 TB+ under treatment.

**Sample Preparation**

A total of 5 mL sputum sample was collected from symptomatic participants. We ran both smear microscopy and standard culture test for labeling the acquired sputum samples. The samples were prepared for culture as described previously [25-27]. Panta, MGIT tube, and growth supplement were used as reagents. Briefly, a mixture was prepared by mixing 15 mL of MGIT growth supplement in Panta, followed by adding 0.8 mL of the resultant enrichment and 0.5 mL of the concentrated specimen to the MGIT tube. Before each step, the MGIT tube was opened for a short time to maintain $O_2$ concentration in the media. The tube was tightly recapped and shaken to blend the solution. The tube and its cap were wiped with mycobactericidal disinfection and placed at room temperature for 30 minutes, followed by placing the tube in the desired MGIT 960 drawer. The tube was monitored daily for indicator lights flagging positive or negative culture test results.

Based on biological test results, samples were categorized into three groups: TB-, first-time TB+, TB+ under treatment. Samples were then filtered through a 0.5-micron filter (Millipore-CA) to remove Mtb bacteria. A 3 mL of cell-free growth supernatant was pipetted into a clean cuvette. The cuvette was tightly closed with the cap to avoid contamination during Raman spectroscopy.

3. 4. **Experimental Setup and Data Post-Processing**

## 4.1 Experimental Setup

A schematic diagram and working principle of the proposed sensor are shown in Figure 1. A continuous-wave laser FPV785S (Thorlabs, Inc.) of optical power 350 mW and wavelength 785 nm is employed as an excitation source. The laser beam is passed through a Thorlabs collimator PAF-X-18-PC-B and then propagated through a narrowband laser line filter to block the unwanted laser modes and thus enhance Raman peaks resolution. The narrow bandwidth laser light is further guided through a long pass dichroic beam splitter (Princeton Instrument, Inc., USA) with a 127 cm-1 edge. The beam splitter reflects the 785 nm and lower wavelengths towards the sample and transmits higher wavelengths. A focusing lens focuses the beam on the sample (a quartz cuvette containing cell-free sputum supernatant), which scatters laser light into Raman stoke (S), antistoke (AS), and Rayleigh (R) light. The scattered light is then back collected and collimated by the focusing lens. The light is propagated through the dichroic beam splitter, which filters out AS and R signals and transmits the S beam. An OD 6 long-pass matching edge filter further enhances AS and R blockage while allowing the S beam to pass. The clean S beam, i.e., our desired Raman signal, is then focused on the entrance slit installed on a diffraction-limited and aberration-free spectrometer (FER-SCI-1024BRX, Princeton, Inc., USA) using an f/4 achromatic doublet focusing lens (NIR 650 -1050 nm). The spectrometer incorporates a cooled CCD detector of size 1024x256 pixels, an entrance slit of width 25 μm, and a grating with grooves density of 1200 grooves/mm. It offers a spectral resolution of 0.13 nm/pixel at a focal length of 8.08 cm and an aperture ratio of f/4. Inside the spectrometer, the light is first collimated through collimation optics, then diffracted into constituent wavelengths using a diffraction grating, and focused on the CCD camera employing a focusing lens. The CCD measures the incident light and consequently records the Raman spectrum.

The probe acquires Raman spectra of sputum's supernatant samples. However, before recording Raman spectra of actual samples, the spectrometer intensity and wavelength were calibrated to minimize the uncertainties induced due to intensity fluctuations and wavelength off-sets. The spectrometer's wavelength off-sets were compensated by adjusting the grating position with the reference Hg-spectrum of the atomic emission light source (Hg/Ne atomic emission light source, Princeton Instruments, Inc., USA) while performing a broadband wavelength calibration over a wavelength range of 200 nm to 1050 nm. Moreover, the spectrometer's intensity was calibrated with a reference intensity calibration lamp (Quartz-Tungsten-Halogen calibration lamp, Princeton Instruments, Inc., USA) to correct the uneven response of CCD pixels.

After the spectrometer calibration, a reference background spectrum was acquired, which was subtracted from the sample's Raman spectrum to cancel out the background environmental effects. In

Raman spectroscopy, the integration time and spectrum averaging are crucial to producing better and reproducible results. The rule of thumb is to select a reasonable integration time, e.g., a duration in which enough Raman signal is gathered, and then average-out multiple Raman spectra to minimize the noise. In our experiments, we acquired the Raman data of each sample by averaging five frames of the corresponding Raman spectrum. The integration time of each Raman spectrum was 1.5 minutes. We also found that the Raman spectrum for the supernatant was highly reproducible. We then performed data post-processing to classify TB infected and non-infected patients. The probe's performance was validated against sputum culture test results. A portion of the unfiltered sputum supernatant was further processed for the TB culture test.

## 4.2 Data Post-Processing

We can represent our measurement matrix, X, as:

$$X = \begin{bmatrix} x_{11} & x_{12} & \cdots & x_{1n} \\ x_{21} & x_{22} & \cdots & x_{2n} \\ \vdots & \vdots & \ddots & \vdots \\ x_{m1} & x_{m2} & \cdots & x_{mn} \end{bmatrix},$$

where $m$ represents the number of collected Raman spectra and $n$ represents data points in each spectrum. In our case, $m = 112$ and $n = 1024$, i.e., 112 Raman spectrums are collected, and each spectrum has 1024 data points. For the classification of samples, we used Principal Component Analysis (PCA) [28]. PCA has been widely utilized for various problems, including image compression and pattern recognition. It has also been applied for analyzing Raman spectrums to find specific patterns, such as the separation of Raman data for cancerous and non-cancerous patients. Mathematically, the data set, X, in PCA is transformed into a new coordinate system using an orthogonal linear transformation. In this new transformation, data is then represented along with the PCA components (eigenvectors) with maximum variation (eigenvalues). The maximum variance is along the first principal component, PC1, the second maximum variance is along the second principal component, PC2, and so on.

In PCA analysis, we first regularized each obtained spectrum from our sensor using the following equation:

$$X_r = \frac{x_i - \mu}{\sigma}$$

Here $x_i$ is an intensity data point, while $\mu$ and $\sigma$ are the mean and standard deviations of all intensity points in the spectrum, respectively. Afterward, we determined the covariance among each data point of the sample's spectrum data. The number of data points in each sample was 1024. Therefore, we obtained a 1024x1024 covariance matrix for every sample. Then we determined the eigenvectors and corresponding eigenvalues of each covariance matrix. The two eigenvectors with maximum eigenvalues represented directions of maximum variance and were named PC1 (principal component 1) and PC2 (principal component 2). Then each spectrum was represented as a data point with the corresponding contributions of PC1 and PC2, as illustrated in Figure 3.

## 5. Results and Discussion

Raman spectra of cell-free supernatant of sputum samples are shown in Figure 2. In Figure 2, the spectrums shown for TB-, First-time TB+, and TB+ under treatment are averages of 45, 18, and 49 collected spectrums, respectively. There are eight prominent and visually coinciding Raman peaks. These peaks correspond to the Raman signature of the sample's supernatant compounds, as indicated by peaks labels. The cell-free supernatant contains MGIT growth supplement, panta, and MGIT tube contents [29]. MGIT growth supplement comprises oleic acid, albumin, catalase, and dextrose. Panta contains azlocillin, amphotericin B, nalidixic acid, trimethoprim, and polymyxin B. Similarly, the MGIT tube comprises modified Middlebrook 7H9 broth media and fluorescence material. These compounds produce specific Raman peaks. Raman peak at 366.3 corresponds to dextrose [30], 552.11 to trimethoprim [31], 597.64 to albumin [32], 810.34 to trimethoprim, 985.66 to oleic acid [33], 1164.09 to catalase [34], 1449.45 to Nalidixic acid [35], and 1724.52 to nalidixic acid, respectively.

However, there are minute changes in the representative Raman spectra of samples due to the presence of TB biomarker compounds even after bacteria filtration. These variations include intensity fluctuation and wavenumber shift of coinciding peaks, as shown in figure 2, leading to the PCA classification. Apart from the noticeable change in peak intensities, we also notice wavelength shifts in Raman peaks. Table 1 presents prominent peak positions of TB-, first-time TB+, and TB+ under treatment for the compounds present in the supernatant.

The classification of 112 TB patients, out of which 45 are TB-, 18 are first-time TB+, and 49 are TB+ under treatment, are shown in Figure 3. The classifier uses a machine-learning algorithm, PCA, on the acquired Raman data to make the diagnosis/classification process robust, rapid, accurate, and technician-free. The sensor judiciously classifies the three groups in the plane of the orthogonal components, e.g., PC1 and PC2. Red, green, and black dots represent TB-, first-time TB+, and TB+ under TB treatment samples, respectively. The scree plot is provided in Figure 4 for the first ten PCs out of 112 PCs. The elbow-point shows that the maximum variation is among the first two principal components and hence retained for the data transformation. The percentage of variance is less than one percent after PC5.

We adopt a linear classifier, where the blue line shows the separation between the three groups. We can see that all positive samples (red and black dots) lie under the blue line and are correctly identified as TB+ with 100% true-positive accuracy. However, 3 TB- samples are misclassified as TB+, bringing down true-negative accuracy to 93.4%. Here, it is worth mentioning that true-negative accuracy is not as critical as true-positive, which can potentially cause TB outspread. Moreover, misclassified samples lie in the proximity of the classification line, indicating the classifier's reliability.

TB- samples are spread over the upper half of the classification plane, while TB+ samples form two clusters in the lower half-plane. There is a separate cluster of TB patients on medication, compactly grouped in the lower half-plane, indicating that the sensor correctly identifies patients on medicines but still with positive TB. However, the cluster of first-time TB+ patients is slightly scattered on the right side in the lower half-plane.

In Table 2, we provide a confusion matrix of PCA of measured Raman spectra of each class. We also provide various performance measures, including accuracy, precision, recall, and F-measure [36] of the classification process in Table 3.

## 4. Conclusion

In this work, we demonstrate for the first time the application of Raman spectroscopy toward rapid TB detection using sputum samples. The sensor offers 100% true-positive and 93.3% true-negative accuracy after testing 112 TB patients. Significantly, our probe can provide a quick (few minutes) TB test result of an already processed sputum sample for the culture test. Therefore, as shown in Figure 1, the probe can be easily integrated into conventional TB diagnostic labs where smear microscopy and culture tests are routinely conducted. One limitation of our work is the requirement of an expensive CCD for realizing the setup. However, the costly sensing platform is potentially replaceable by our recently demonstrated optical fiber sensing systems [37]. We anticipate that the demonstrated probe will lead to a viable, cost-effective, accurate, and rapid test for TB and other infectious diseases, including malaria and pneumonia. Consequently, the diagnostic landscape of infectious diseases can be immensely impacted, especially in low-resource countries.

**Funding**

This work was funded by IGNITE-National Technology Fund, Pakistan, through grant no. ICTRDF/TR&D/2016/42.

**Declaration of competing interest**

The authors declare that they have no conflict of interest.

**Acknowledgment**

We thank Asma Idress, Farrah Naz, and Habib-ur-Rehman (medical staff at the Institute of Public Health) for their assistance in providing and validating human samples.

| Compounds | TB $-$ ($cm^{-1}$) | First-time TB+ ($cm^{-1}$) | TB+ under treatment ($cm^{-1}$) |
|---|---|---|---|
| Dextrose | 366.72 | 366.72 | 368.8 |
| Albumin | 597.58 | 600.08 | 598.41 |
| Trimethoprim | 809.64 | 811.18 | 811.39 |
| Oleic Acid | 984.47 | 984.08 | 985.95 |
| Catalase | 1164.1 | 1165.7 | 1166.67 |
| Nalidixic Acid | 1449 | 1449.5 | 1450.55 |
| Nalidixic Acid | 1722.6 | 1722.9 | 1726.35 |

Table 1: Prominent peak positions of TB-, first-time TB+, and TB+ under treatment for the compounds present in the supernatant

| Total 18+45+49=112 | Predicted TB+ | Predicted TB- | Predicted TB+ under Treatment |
|---|---|---|---|
| Actual TB+ (18) | 18 | 0 | 0 |
| Actual TB- (45) | 2 | 42 | 1 |
| Actual TB+ Treatment (49) | 0 | 0 | 49 |

Table 2: Confusion matrix of PCA of measured data

| Class | Accuracy | Precision | Recall | F-Measure |
|---|---|---|---|---|
| TB+ | 0.97 | 0.86 | 1 | 0.92 |
| TB- | 0.97 | 1 | 0.93 | 0.97 |
| TB+ under Treatment | 0.97 | 0.94 | 1 | 0.97 |

Table 3: Performance of the classification on various measures

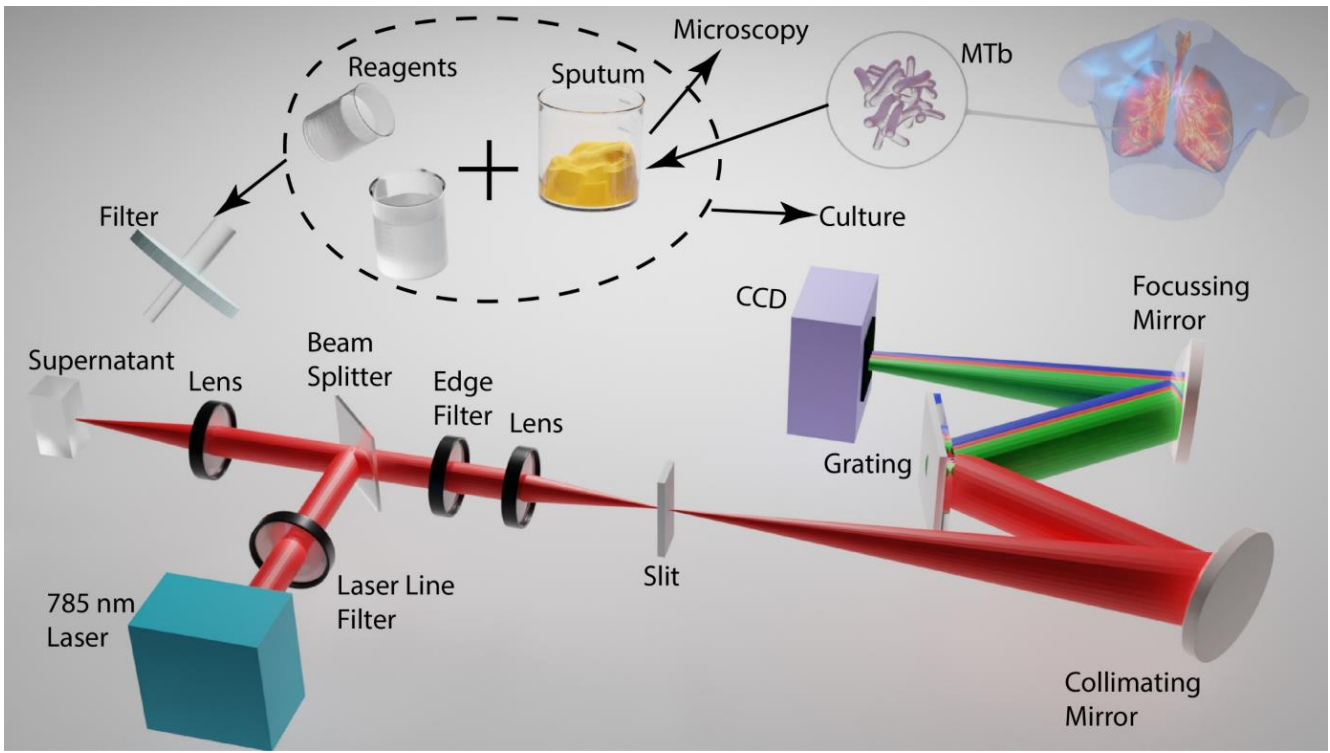

Figure 1. Schematic diagram of the sensor. (Please note that the beam does not have any true colors; rather, it is just for concept illustration).

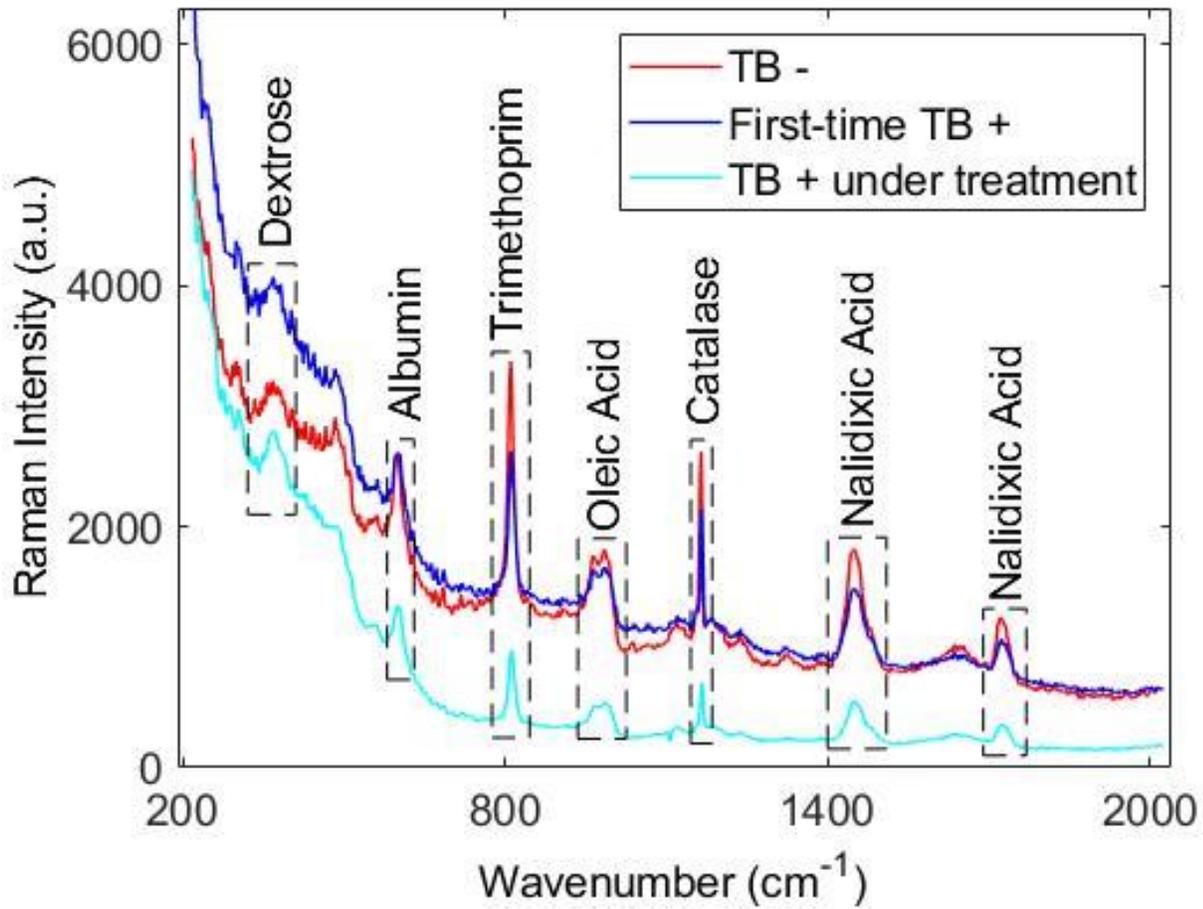

Figure 2. Average Raman spectra of the three classification groups. The spectrums shown for TB-, First-time TB+, and TB+ under treatment are averages of 45, 18, and 49 collected spectrums, respectively. The dashed rectangles indicate Raman signature peaks of cell-free sputum samples. We label each peak with its corresponding compound in the sample's supernatant.

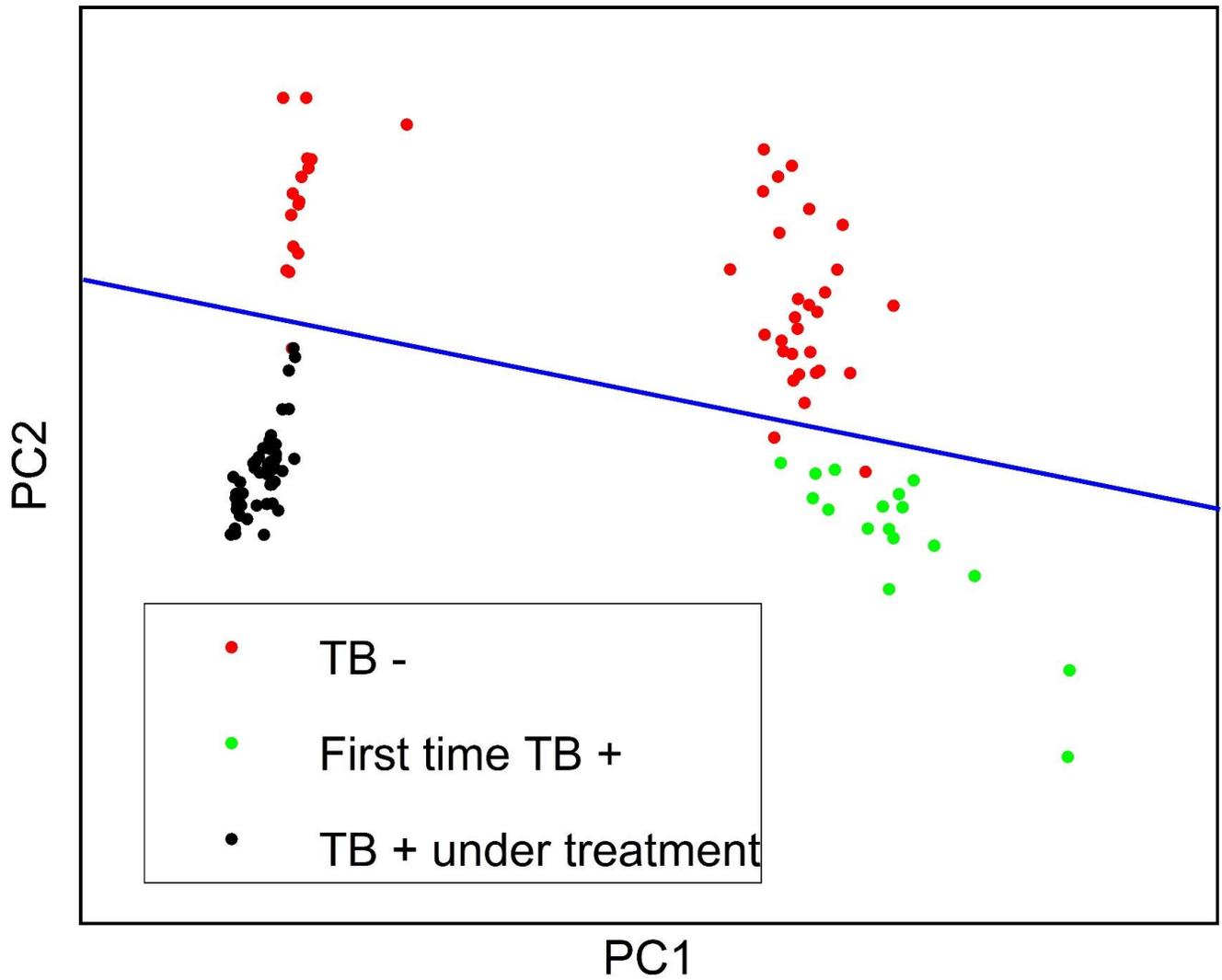

Figure 3. TB diagnostic results in under seven minutes with the portable optical probe. Red dots represent TB-negative, green dots denote recently diagnosed TB-positive, and black dots show TB-positive receiving TB treatment samples.

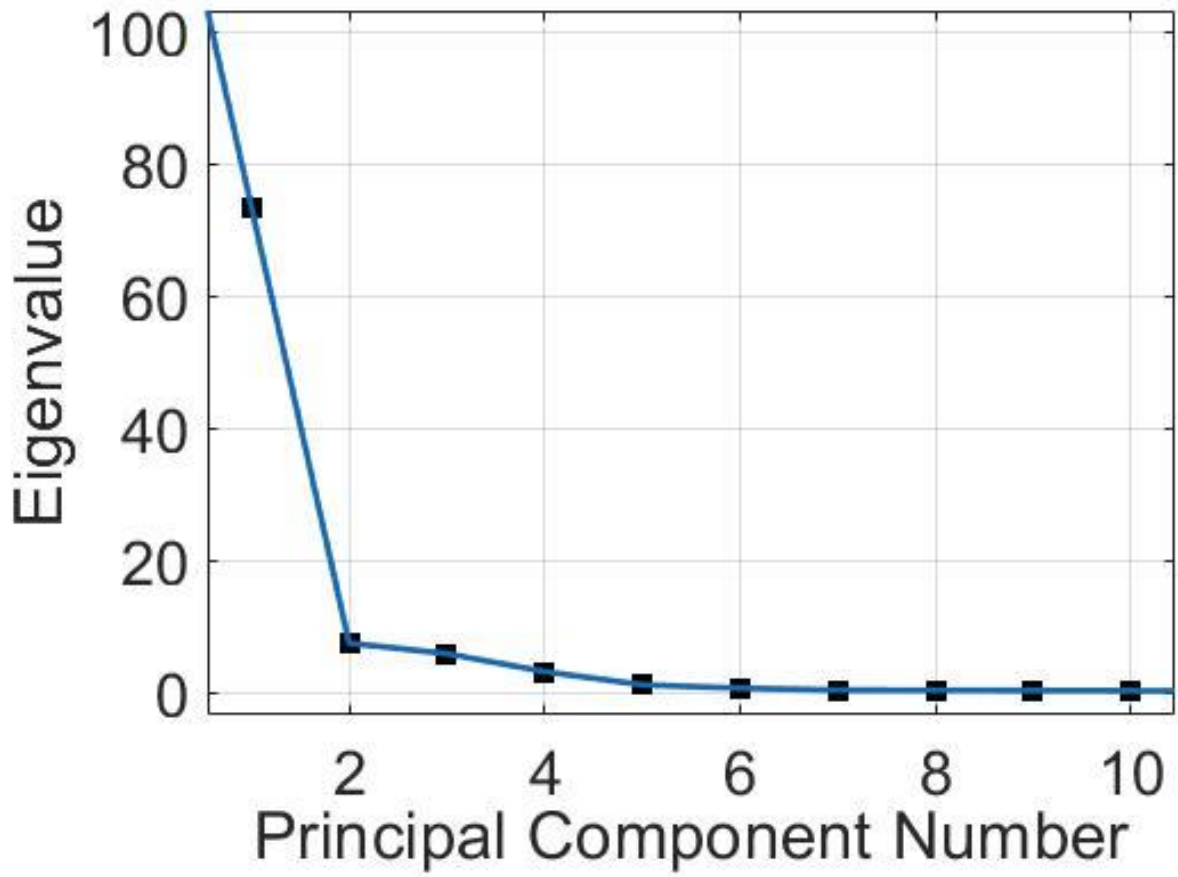

Figure 4. Scree plot for the first ten principal components.